\documentclass{ifacconf}

\usepackage{graphicx}      
\usepackage[sort&compress]{natbib}        
\usepackage{url}

\newtheorem{theorem}{Theorem}
\newtheorem{lemma}{Lemma}

\newtheorem{remark}{Remark}

\newtheorem{assumption}{Assumption}

\usepackage{amsmath} 
\usepackage{amssymb}  
\usepackage{mathtools}
\usepackage{xcolor}
\usepackage{customCommands}

\usepackage{enumitem}
\newlist{CustomList}{enumerate}{1}
\setlist[CustomList,1]{label={A\arabic*}),ref={A\arabic*}}

\newcounter{constnum}

\usepackage{algorithm}
\usepackage{algpseudocode}

\usepackage[normalem]{ulem}

\begin{document}
\begin{frontmatter} 

\title{
Non-Asymptotic Error Bounds for Causally Conditioned Directed Information Rates of Gaussian Sequences\thanksref{footnoteinfo}
}

\thanks[footnoteinfo]{
This work was supported in part by NSF ECCS 2412435 and NIH 1R01NS147767-01.  
}

\author[First]{Yuping Zheng} 
\author[First]{Andrew Lamperski} 

\address[First]{University of Minnesota, Twin Cities, Minneapolis, MN 55455 (e-mail: zhen0348@umn.edu, alampers@umn.edu)}



\begin{abstract}
Directed information and its causally conditioned variations are often used to measure causal influences between random processes. In practice, these quantities must be measured from data. Non-asymptotic error bounds for these estimates are known for sequences over finite alphabets, but less is known for real-valued data. This paper examines the case in which the data are sequences of Gaussian vectors. 
We provide an explicit formula for causally conditioned directed information rate based on optimal prediction and define an estimator based on this formula. We show that our estimator gives an error of order $O\left(N^{-1/2}\log(N)\right)$ with high probability, where $N$  is the total sample size. 
\end{abstract}

\begin{keyword}
  Directed Information, Causality, Optimal Prediction
\end{keyword}

\end{frontmatter}
\section{INTRODUCTION}

Directed information (DI) measures the direction of the information flow in  coupled dynamical systems. 
Over the years, directed information has been widely used for diverse areas including networked control systems \citep{tanaka2017lqg,tanaka2017directed}, econometrics \citep{permuter2011interpretations,etesami2017econometric}, neuroscience \citep{quinn2011estimating,tsur2023neural}, genetics \citep{rao2007using,rao2007motif}, and social sciences \citep{ver2012information}.

The idea of directed information originated from the work of \citep{marko1973bidirectional} in the context of bidirectional communication theory. Based on these ideas, \citep{massey1990causality} introduced the definition of directed information for a communication channel with feedback. 
Causally conditioned directed information was introduced in \citep{kramer1998directed}, and can be viewed as generalization of directed information.

Another information measure closely related to DI, and commonly studied in physics, is transfer entropy \citep{geweke1982measurement, schreiber2000measuring,barnett2009granger, hempel2024simple}. 


The review \citep{amblard2012relation} points out that tools from directed information theory can be naturally used to measure  Granger causality \citep{granger1969investigating}. In particular,  it is shown in \citep{amblard2011directed} that  DI theory can be used to study Granger causality graphs of time series. In particular, linear Granger causality and DI are equivalent for jointly Gaussian processes \citep{quinn2015directed,hempel2024simple,amblard2012relation}.

There have been extensive studies on using directed information to determine the structure of networked stochastic dynamic systems \citep{quinn2011equivalence,quinn2015directed,quinn2017bounded,young2021inferring,subramanian2021effects,etesami2014directed,etesami2017econometric,wang2018relationship,divernois2024analysis,etesami2023modeling,hempel2024simple}.  




For sequences over finite alphabets, non-asymptotic error bounds for directed information and causally conditioned variants were given in \citep{jiao2013universal,quinn2015directed}. However, no non-asymptotic bounds are known for data from infinite sets. 

This paper examines the special case of data from infinite sets in which the data are sequences of Gaussian vectors. The main contribution is an estimation scheme for causally conditioned directed information rate which achieves an error of order $O(N^{-1/2}\log(N))$, where $N$ is the length of the time series data. To derive the estimator, we provide an analytic formula  for the causally conditioned directed information rate based on optimal prediction theory. 
  Several closely related formulas appear in prior works \citep{etesami2014directed,divernois2024analysis,barnett2009granger,chicharro2011spectral,hempel2024simple} and match  intermediate steps of the derivation of Theorem~\ref{thm:exact}. However, none of them obtain our specific formula for causally conditioned DI rate, nor do they provide non-asymptotic error bounds.

The rest of the paper is organized as follows. In Section \ref{sec:Setup}, we present preliminaries to define the problem. Section~\ref{sec:results} gives the main results. The analytic formula for causally conditioned directed information rate is proved in Section~\ref{sec:exactProof} and the error bound for the estimator is proved in Section~\ref{sec:estimatorProof}. Section~\ref{sec:simulation} presents a simple simulation to verify the theoretical result. Conclusions are given in Section~\ref{sec:conclusion}.

\section{Problem Setup} \label{sec:Setup}

\subsection{Notation}
%
%
%
$\bbR$, $\bbC$ denote the sets of real and complex numbers.
$I$ denotes identity matrix. 
$\|x\|_2$ denotes the Euclidean norm for vectors and the spectral norm for matrices. $\|A\|_F$ denotes the Frobenius norm of matrix.
$A^\star$ denotes the conjugate transpose while $A^\top$ denotes the transpose. We use Python-style indexing for vectors: $y[i:k]$ denotes the vertical stack $\begin{bmatrix}y[i]^\top & \cdots & y[k-1]^\top\end{bmatrix}^\top$, which is empty if $k\le i$. 
For Hermitian matrices $A$ and $B$, $A \succ B (A \succeq B)$ denotes that $A-B$ is positive definite (semidefinite). 
Trace of a square matrix $A$ is denoted by $\Tr(A)$ and the determinant is $\det(A)$.
Bold letters denote random variables and the expectation of a random variable, $\by$, is denoted by $\bbE[\by]$. $\bbP$ denotes probability.



The unit disc of complex number is denoted by $\bbD= \left\{z \in \bbC: |z|<1\right\}$ and its closure is denoted by $\bar{\bbD}$. The unit circle in $\bbC$ is denoted by $\bbT$.
If $M:\bbT \to \bbC^{m\times n}$ is a matrix-valued function, then $\|M\|_{L_{\infty}}=\esssup_{z\in\bbT}\|M(z)\|_2$.

\subsection{Causally Conditioned Directed Information Rates}
If $\bx$, $\by$, and $\bz$ are random variables with joint density $p_{XYZ}$ with respect to Lebesgue measure, then the mutual information of $\bx$ and $\by$, conditioned on $\bz$ is given by:
\begin{align}
  I(\bx;\by|\bz)
  \label{eq:conditionalMIBayes}
                &=\bbE\left[\log\frac{p_{Y|XZ}(\by|\bx,\bz)}{p_{Y|Z}(\by|\bz)}\right].
\end{align}
Here  $p_{Y|Z}$ is  the conditional density of $\by$ given $\bz$ while $p_{Y|XZ}$ is the conditional density of $\by$ given $(\bx,\bz)$. 

For stochastic processes $\bx$, $\by$, $\bz$,  causally conditioned directed information is defined as in \citep{quinn2015directed} by:
\begin{multline}
  \label{eq:ccdi}
I(\bx[0:k]\to \by[0:k]\| \bz[0:k])=\\\sum_{i=1}^{k-1}I(\bx[0:i];\by[i] | \by[0:i],\bz[0:i]).
\end{multline}
When $\bx$, $\by$, and $\bz$ are vector-valued, our Python-style indexing convention gives that 
$$\bx[0:i]=\begin{bmatrix}\bx[0]^\top & \cdots & \bx[i-1]^\top\end{bmatrix}^\top.
$$ 
We do not include the term with $i=0$ in the sum, since, in this case, $\bx[0:0]$, $\by[0:0]$, and $\bz[0:0]$ correspond to empty lists of vectors.  

The definition of causally conditioned directed information from \eqref{eq:ccdi} corresponds to Equation (5) from \citep{quinn2015directed}, and quantifies how much information the strict past values  $\bx[0],\ldots,\bx[i-1]$ contains about the present value of $\by[i]$, conditioned on the strict past values $\by[0],\ldots,\by[i-1]$ and $\bz[0],\ldots, \bz[i-1]$. The original definition from  \citep{kramer1998directed} deals with non-strict conditioning, and in that work, the terms in the sum actually correspond to $I(\bx[0:i+1];\by[i]|\by[0:i],\bz[0:i+1])$ in our notation.

The causally conditioned directed information rate is 
$$
I^{\infty}(\bx\to\by \| \bz) = \lim_{k\to\infty}\frac{1}{k} I(\bx[0:k]\to \by[0:k]\| \bz[0:k]).
$$

\subsection{Problem}

Let $\bw[k]=\begin{bmatrix}\bx[k]^\top & \by[k]^\top & \bz[k]^\top \end{bmatrix}^\top$ be a stationary stochastic process such that each $\bw[k]$ is a zero mean vector with real entries. The autocovariance sequence and power spectral density (PSD) are
$$R[k] = \bbE[\bw[i+k]\bw[i]^\top]
\quad
\Phi(z) = \sum_{k=-\infty }^{\infty} z^{-k} R[k].
$$ 

We make the following assumptions on $\bw[k]$:
\begin{assumption}
  There exist positive $c_{\min}$ and $c_{\max}$, such that
$c_{\min} I \preceq \Phi(e^{j \omega}) \preceq c_{\max}I$ for all $\omega \in[0, 2\pi]$. \label{assump:postive_psd}
\end{assumption}
\begin{assumption} \label{assump:rational_PSD}$\Phi(z)$ is rational. 
\end{assumption}
\begin{assumption}\label{assump:gaussian} $\bw[k]$ is Gaussian.
\end{assumption}
\begin{remark}
Assumption of positive rational spectral density guarantees the existence of the finite-dimensional state-space realization and further the existence of the optimal predictor. The upper bound assumption on the spectral density ensures the system is well-behaved across all frequencies. Assumption \ref{assump:gaussian} enables the simple formula for the exact causally conditioned DI rate calculation.
\end{remark}
Under these assumptions, the problem studied in this paper  is to compute estimates of the causally conditioned directed information rate, $I^{\infty}(\bx\to\by \| \bz)$, from a finite collection of data, $\bw[0:N]$, and quantify the error.

\section{Information Rate Formula and Estimate}
\label{sec:results}

\subsection{Optimal Prediction}
Recall that $\bw[k]=\begin{bmatrix}\bx[k]^\top & \by[k]^\top & \bz[k]^\top\end{bmatrix}^\top$. Set $\bv[k]=\begin{bmatrix}\by[k]^\top & \bz[k]^\top \end{bmatrix}^\top$. Then $\bv$ satisfies Assumptions~\ref{assump:postive_psd}, \ref{assump:rational_PSD}, and \ref{assump:gaussian}, with the same constants $c_{\min}$ and $c_{\max}$ for Assumption~\ref{assump:postive_psd}.

Assumptions~\ref{assump:postive_psd} and ~\ref{assump:rational_PSD} imply that there is a strictly causal transfer matrices, $H$ and $J$, such that the minimum mean-squared error predictors for $\bw[k]$ and $\bv[k]$ are given by:
\begin{align*}
  \hat \bw[k]&=\sum_{i=1}^{\infty}H_i \bw[k-i] 
             &\hat \bv[k]&=\sum_{i=1}^{\infty}J_i \bv[k-i].
\end{align*}
The predictions, $\hat\bw[k]$ and prediction errors $\bw[k]-\hat\bw[k]$ are zero-mean by the assumption that $\bw[k]$ are zero mean. 
Under Assumptions~\ref{assump:postive_psd} and \ref{assump:rational_PSD}, the prediction errors $\bw[k]-\hat\bw[k]$ are orthogonal of the linear span of the past data. Specifically, if $\bzeta$ is vector which is a linear function of $\bw[k-1],\bw[k-2],\ldots$, then $\bbE[(\bw[k]-\hat\bw[k])\bzeta^\top]=0$. See \cite{wiener1957prediction}. 

Similarly, $\hat\bv[k]$ gives the minimum mean-squared error prediction of $\bv[k]$, given $\bv[k-1],\bv[k-2],\ldots$, and the prediction errors are zero mean and orthogonal to linear functions of the past measurements.

Under Assumption~\ref{assump:rational_PSD}, $H$ and $J$ can be computed from the Kalman filter, if $\Phi$ were known. See~\cite{kailath2000linear}.

Denote the prediction error covariance matrices by:
\begin{subequations}
  \label{eq:errorCov}
\begin{align}
  \Sigma&=\begin{bmatrix}\Sigma_{xx} & \Sigma_{xy} & \Sigma_{xz} \\
    \Sigma_{yx} & \Sigma_{yy} & \Sigma_{yz} \\\
    \Sigma_{zx} & \Sigma_{zy} & \Sigma_{zz}
  \end{bmatrix} = \bbE\left[(\bw[k]-\hat\bw[k]) (\bw[k]-\hat\bw[k])^\top\right] \\
    \Gamma&=\begin{bmatrix}\Gamma_{yy} & \Gamma_{yz} \\
    \Gamma_{zy} & \Gamma_{zz}\end{bmatrix}= 
\bbE\left[(\bv[k]-\hat\bv[k]) (\bv[k]-\hat\bv[k])^\top\right].
\end{align}
\end{subequations}

\subsection{Causally Conditioned Directed Information Rates from Optimal Predictors}
 
The following theorem gives an exact formula for the causally conditioned directed information rate in terms of the prediction error covariances described above. It is proved in Section~\ref{sec:exactProof}.

\begin{theorem}
  \label{thm:exact}
  {\it
  Let $\bw[k]=\begin{bmatrix}\bx[k]^\top & \by[k]^\top &\bz[k]^\top \end{bmatrix}$ be a stationary, zero-mean stochastic process satisfying Assumptions~\ref{assump:postive_psd}, \ref{assump:rational_PSD}, and \ref{assump:gaussian}. Let $\Sigma$ and $\Gamma$ be the matrices from \eqref{eq:errorCov}. The causally conditioned directed information rate is given by:
  \begin{align*}
  I^{\infty}(\bx\to\by\| \bz)
  =\frac{1}{2}\log\frac{\det(\Gamma_{yy})}{\det(\Sigma_{yy})}.
\end{align*}
}
\end{theorem}

\subsection{Estimating Causally Conditioned DI Rates}

Here we present our estimation procedure and corresponding error bound. The estimation procedure is based on the Theorem~\ref{thm:exact} and the following preliminary result:
\begin{lemma}
  \label{lem:covConvergence}
  {\it 
    For $i\ge 1$ let $A_i = \bbE[\bw[0:i]\bw[0:i]^\top]$ and $B_i = \bbE[\bw[i]\bw[0:i]^\top]$. Let $\Sigma$ be the error covariance matrix from \eqref{eq:errorCov}. Let $H$ be the optimal predictor for $\bw$. Let $\rho \in (0,1)$ be such that every  pole, $s$, of $H$, satisfies $|s|<\rho$ and let $b\ge \sup_{|z|\ge \rho}\|H(z)\|_2$. 

If Assumptions~\ref{assump:postive_psd} and ~\ref{assump:rational_PSD} hold,  then for all $i\ge 1$ the following bound holds:
  $$
  \|\Sigma- \left(R[0]-B_i A_i^{-1}B_i^\top \right)\|_2 \le \frac{b^2 \rho^{2(i+1)}}{(1-\rho)^2}c_{\max}.
  $$
}
\end{lemma}
An analogous bound based on the properties of the predictor, $J$ also holds. To cover both cases, let $\rho\in (0,1)$ be such that $|s| \le \rho$ if $s$ is a pole of either $H$ or $J$, and let $b=\max\{\sup_{|z|\ge \rho}\|H(z)\|_2,\sup_{|z|\ge \rho}\|J(z)\|_2\}$.

For $p\ge 1$, note that 
$$
\bbE[\bw[0:p+1]\bw[0:p+1]^\top]= \begin{bmatrix}A_p & B_p^\top \\
  B_p & R[0]
\end{bmatrix}.
$$
So, Lemma~\ref{lem:covConvergence} indicates that to estimate $\Sigma$, it suffices to estimate $\underline{R}:=\bbE[\bw[0:p+1]\bw[0:p+1]^\top]$ for sufficiently large $p$. Similarly, to estimate $\Gamma$, it suffices to estimate $\underline{Q}:=\bbE[\bv[0:p+1]\bv[0:p+1]^\top]$. 

Let $\bw[0:N]$ be the available data set. (Recall that $\bv[i]$ is a subvector of $\bw[i]$.) Fix $1\le p < N$ and set $M=N-p$. Estimates of $\underline{R}$ and $\underline{Q}$ are given by
\begin{subequations}
  \label{eq:empCov}
\begin{align}
  \tilde\bR&=\frac{1}{M}\sum_{k=0}^{M-1}\bw[k:k+p+1] \bw[k:k+p+1]^\top\\
  \tilde\bQ&=\frac{1}{M}\sum_{k=0}^{M-1}\bv[k:k+p+1]\bv[k:k+p+1]^\top.
\end{align}
\end{subequations}

Partition the estimates as:
\begin{align*}
  \tilde\bR&=\begin{bmatrix}
    \tilde\bA & \tilde\bB^\top \\
    \tilde\bB & \tilde \bR[0]
  \end{bmatrix} 
           &
  \tilde\bQ&=\begin{bmatrix}
    \tilde\bC & \tilde\bD^\top \\
    \tilde\bD & \tilde \bQ[0]
  \end{bmatrix}.
\end{align*}
Then our estimates for the error covariances, $\Sigma$ and $\Gamma$ are given, respectively by:
\begin{align*}
  \tilde\bSigma&=\tilde\bR[0]-\tilde\bB\tilde\bA^{-1}\tilde\bB^\top=
  \begin{bmatrix}
    \tilde\bSigma_{xx} & \tilde\bSigma_{xy} & \tilde\bSigma_{xz} \\
    \tilde\bSigma_{yx} & \tilde\bSigma_{yy} & \tilde\bSigma_{yz} \\
    \tilde\bSigma_{zx} & \tilde\bSigma_{zy} & \tilde\bSigma_{zz}
  \end{bmatrix}
  \\
  \tilde\bGamma&=\tilde\bQ[0]-\tilde\bD\tilde\bC^{-1}\tilde\bD^\top=
  \begin{bmatrix}
    \tilde\bGamma_{yy} & \tilde\bGamma_{yz} \\
    \tilde\bGamma_{zy} & \tilde\bGamma_{zz}
  \end{bmatrix}.
\end{align*}
In our analysis, we will show that the required inverses exist with high probability, for sufficiently large $M$. 

Our estimate for causally conditioned directed information rate is then given by: 
$$
\tilde\bI = \frac{1}{2}\log\frac{\det(\tilde\bGamma_{yy})}{\det(\tilde\bSigma_{yy})}.
$$

Our main result is the following non-asymptotic bound. It is proved in Section~\ref{sec:estimatorProof}.

\begin{theorem}
  \label{thm:error}
  {\it Let $N$, $M$, $p$, $I(\bx\to\by\| \bz)$, and $\tilde\bI$ be defined as above. Say that $\by \in\bbR^{n_Y}$, $\bw \in \bbR^{n_W}$. For $\nu \in (0,1)$, set 
  \begin{align*}
    \MoveEqLeft[0]
    \epsilon=\\
    &
    2 c_{\max}\max\left\{
  (p+1)8 \left(\frac{\log(2\nu^{-1})}{M} + \frac{n_W(1+p)}{M} \log (13) \right)
    \!, \right.  \\
    & \left.  
\sqrt{(2p+1)8 \left(\frac{\log(2\nu^{-1})}{M} \!+ \!\frac{n_W(1+p)}{M} \log (13) \right)}
    \right\}.
  \end{align*}
  If $\epsilon < c_{\min}$, then with probability at least $1-\nu$, the inverses required to compute $\tilde \bI$ exist, and the following error bound holds:
  \begin{multline*}
|I^\infty(\bx\to\by\| \bz)-\tilde\bI|\le  \frac{n_Y}{c_{\min}-\epsilon}\cdot \\
\left(
c_{\max}\frac{b^2 \rho^{2(p+1)}}{(1-\rho)^2}+
\epsilon\left(1+\frac{(c_{\max}+\epsilon)^2}{(c_{\min}-\epsilon)^2} \right)
\right).
  \end{multline*}

  In particular, if $a_1 \log(N) \le p \le a_2 \log(N)$ for some $a_2\ge a_1 \ge \frac{1}{4\log(\rho^{-1})}$, then with probability at least $1-\nu$
  $$
|I^\infty(\bx\to\by\| \bz)-\tilde\bI|=O\left(\log(N)\sqrt{\frac{\log(\nu^{-1})}{N}}\right) .
  $$
}
\end{theorem}

\begin{remark}
  Recall that $\rho \in (0,1)$ is a bound on the magnitudes of the poles of $H$ and $J$. To obtain the last error bound above, we need to know $\rho$. When estimating the causally conditioned directed information rates from data, such a bound will typically be unknown, and so choosing $p=\Theta(\log(N))$ could result in $\rho^{2p}$ shrinking more slowly than $N^{-1/2}$. This problem could be resolved by choosing a polylogarithmic $p$, e.g., $p=(\log(N))^2$, which does not rely on knowing $\rho$, but results in the slightly worse bound of  $O\left((\log(N))^2\sqrt{\frac{\log(\nu^{-1})}{N}}\right)$. Additionally, the error bound scales with the data dimension at the rate of $O(n_Y n_W^3)$, which indicates that our error bound is loose for high-dimensional data.
\end{remark}

\begin{remark}
  Our estimators for $\Sigma$ and $\Gamma$ are precisely the empirical covariances of the residual errors from order $p$ vector autoregressive models (fit by least-squares) for $\bw$ and $\bv$, respectively. So, the estimates can be computed from autoregessive fits, rather than directly from empirical covariances of $\bw$ and $\bv$. 
\end{remark}

\section{Proof of Theorem~\ref{thm:exact}}
\label{sec:exactProof}

First we present some preliminary facts, including a proof of Lemma~\ref{lem:covConvergence}. Then Theorem~\ref{thm:exact} is proved.

\subsection{Preliminary results}
\begin{lemma}
  \label{lem:covBounds}
  {\it
  If Assumption~\ref{assump:postive_psd} holds, then for all integers with $i < k$, 
  $$
  c_{\min}I\preceq \bbE[\bw[i:k]\bw[i:k]^\top]\preceq c_{\max} I.
  $$
}
\end{lemma}
{\it Proof:}
The following proof is modified from Lemma 3 in \citep{lamperski2023nonasymptotic}.

Let $\underline{R}_w= \bbE[\bw[i:k]\bw[i:k]^\top]$. Assume that $\bw[\ell]\in\bbR^{n_{W}}$ so that $\underline{R}_w$ is an $(k-i)n_W\times (k-i)n_W$ matrix. For any vector, $q\in \bbR^{(k-i)n_W}$, there is a signal $u[\ell]\in\bbR^{n_W}$, such that $q=u[0:k-i]$, $u[\ell]=0$ for $\ell <0$, and $u[\ell]=0$ for $\ell \ge k-i$. Let $\hat u$ be the Fourier transform of $u$. Then using Assumption~\ref{assump:postive_psd}, Plancharel's theorem, and direct computations gives:
\begin{align*}
  c_{\min}q^\top q &= \frac{c_{\min}}{2\pi}\int_{0}^{2\pi}\hat u(e^{j\omega})^\star \hat u(e^{j\omega})d\omega \\
                   &\le \frac{1}{2\pi}\int_{0}^{2\pi}\hat u(e^{j\omega})^\star \Phi(e^{j\omega}) \hat u(e^{j\omega})d\omega \\
                   &=q^\top \underline{R}_w q \\
                   &\le  \frac{c_{\max}}{2\pi}\int_{0}^{2\pi}\hat u(e^{j\omega})^\star  \hat u(e^{j\omega})d\omega \\
                   &=c_{\max}q^\top q.
\end{align*}
\hfill$\qed$

The next result is a variation of Lemma~2 of \cite{lee2020non}, which is, in turn, a generalization of Lemma~1 of \cite{goldenshluger2001nonasymptotic} from scalar-valued transfer functions to transfer matrices. 

\begin{lemma}
  \label{lem:predictorCoeffBound}
  {\it
    Let $\Phi$ satisfy Assumptions \ref{assump:postive_psd} and \ref{assump:rational_PSD}. Let $H$ be the corresponding optimal predictor. Let  $\rho \in (0,1)$ be such that every pole, $s$, of  $H$ satisfies $|s| < \rho$. Then $\sup_{|z|\ge \rho}\|H(z)\|_2 <\infty$. If $H_{i,\mathrm{tail}}(z)=\sum_{k=i+1}^{\infty}H_kz^{-k}$ for $i\ge 1$ and $b\ge \sup_{|z|\ge \rho}\|H(z)\|_2$, then
    $$
    \|H_{i,\mathrm{tail}}\|_{L_{\infty}}\le \frac{b \rho^{i+1}}{1-\rho}.
    $$
}
\end{lemma}
{\it Proof (summary):}
From the assumptions, we know that $\|H(z^{-1})\|_2 \le b$ for all $|z| \le \rho^{-1}$. Furthermore, using Cauchy's estimate gives $\|H_i\|_2 \le b \rho^i$. Then, applying geometric-series bound completes the proof.
\hfill $\qed$

The following elementary fact will be used in a few places, including Lemma~\ref{lem:predictionCovBounds}, which follows afterwards.
\begin{lemma}
  \label{lem:completeSquare}
  {\it
  Let $\begin{bmatrix}
  A& B^\top \\
  B & C
  \end{bmatrix}$ be a block matrix such that $A$ is positive definite. For any $\Theta$ such that $\Theta B^\top$ and $C$ have the same dimensions, the following bound holds:
  \begin{align*}
  C-BA^{-1}B^\top &\preceq C-\Theta B^\top - B\Theta^\top +\Theta A \Theta^\top \\
                  &=\begin{bmatrix}
  -\Theta & I
\end{bmatrix}\begin{bmatrix}
  A & B^\top \\
  B & C
\end{bmatrix}
\begin{bmatrix}
-\Theta^\top \\
I
\end{bmatrix}.
  \end{align*}
}
\end{lemma}
{\it Proof:}
  The equality is a direct calculation. The inequality follows from completing the square:
  \begin{multline*}
  C-\Theta B^\top - B\Theta^\top +\Theta A \Theta^\top = \\
  C-BA^{-1}B^\top +(\Theta - BA^{-1})A(\Theta-BA^{-1})^\top.
  \end{multline*}
\hfill$\qed$

\begin{lemma}
  \label{lem:predictionCovBounds}
  {\it
    For $i\ge 1$ let $A_i = \bbE[\bw[0:i]\bw[0:i]^\top]$, $B_i = \bbE[\bw[i]\bw[0:i]^\top]$, and let $H_{i,\mathrm{tail}}$ be as in Lemma~\ref{lem:predictorCoeffBound}. Let $\Sigma$ be the error covariance matrix from \eqref{eq:errorCov}. If Assumptions~\ref{assump:postive_psd} and ~\ref{assump:rational_PSD} hold,  then for all $i\ge 1$ the following bounds hold:
    \begin{align*}
      \Sigma 
             &\preceq R[0]-B_iA_i^{-1} B_i^\top \\
             &\preceq \Sigma +
 \frac{1}{2\pi}\int_{0}^{2\pi}H_{i,\mathrm{tail}}(e^{i\theta})\Phi(e^{j\theta})
                            (H_{i,\mathrm{tail}}(e^{i\theta}))^\star
d\theta.
    \end{align*}
}
\end{lemma}

{\it{Proof:}}
First, we prove a more general bound that implies the first inequality. Recall that the prediction errors, $\bw[i]-\hat\bw[i]$ are orthogonal to the linear span of $\bw[i-1],\bw[i-2],\ldots$. For any $\Theta$ such that $\bw[i]$ and $\Theta\bw[0:i]$, have the same dimensions, we have:
\begin{align}
  \MoveEqLeft[1]
  \nonumber
  R[0]-\Theta B_i^\top -B_i \Theta^\top +\Theta A_i \Theta^\top  \\
  \nonumber
  &=
  \bbE[\left(\bw[i]-\Theta \bw[0:i]\right)
\left(\bw[i]-\Theta \bw[0:i]\right)^\top
  ] \\
  \nonumber
  &= \bbE\left[\left(\bw[i]-\hat \bw[i]+\hat\bw[i]-\Theta \bw[0:i]\right)\cdot\right. \\
  \nonumber
  &\left.\left(\bw[i]-\hat\bw[i]+\hat\bw[i]-\Theta \bw[0:i]\right)^\top 
  \right] \\
  \label{eq:remainderCov}
  &=\Sigma + \bbE\left[(\hat \bw[i]-\Theta \bw[0:i])(\hat\bw[i]-\Theta\bw[0:i])^\top \right] \\
  \nonumber
  &\succeq \Sigma,
\end{align}
where the third equality follows from the orthogonality property.

Choosing $\Theta = B_iA_i^{-1}$ implies that $\Sigma\preceq R[0]-B_iA_i^{-1}B_i^\top$ for all $i\ge 1$. 

Finally, we prove the second inequality in the lemma.

For general $\Theta$ as above, we have
\begin{align}
  \nonumber
  \MoveEqLeft[1]
  \bbE[(\bw[i]-\Theta \bw[0:i])(\bw[i]-\Theta \bw[0:i])^\top] \\
  \nonumber
  &=R[0]-\Theta B_i^\top -B_i \Theta^\top +\Theta A_i \Theta^\top  \\
  \label{eq:covLB}
  &\succeq R[0]-B_i A_i^{-1}B_i^\top.
\end{align}
The second equality follows from Lemma~\ref{lem:completeSquare}.

Set $\Theta = \begin{bmatrix}H_i & \cdots & H_1\end{bmatrix}$. Then
$$
\hat\bw[i]-\Theta\bw[0:i]=\sum_{k=i+1}^{\infty}H_k \bw[i-k].
$$
In this case, \eqref{eq:covLB} and \eqref{eq:remainderCov} imply that
\begin{align*}
  R[0]-B_i A_i^{-1}B_i^\top &\preceq R[0]-\Theta B_i^\top -B_i\Theta^\top +\Theta A_i \Theta^\top \\
                            &
                            \hspace{-40pt}
                            =\Sigma + \frac{1}{2\pi}\int_{0}^{2\pi}H_{i,\mathrm{tail}}(e^{i\theta})\Phi(e^{j\theta})
                            (H_{i,\mathrm{tail}}(e^{i\theta}))^\star
d\theta.
\end{align*}
\hfill$\qed$

{\it Proof of Lemma~\ref{lem:covConvergence}:}
Lemma~\ref{lem:predictionCovBounds} implies that 
\begin{align*}
  0&\preceq \left(R[0]-B_i A_i^{-1}B_i^\top \right) - \Sigma \\
   &\preceq 
\frac{1}{2\pi}\int_{0}^{2\pi}H_{i,\mathrm{tail}}(e^{i\theta})\Phi(e^{j\theta})
                            (H_{i,\mathrm{tail}}(e^{i\theta}))^\star d\theta.
                          \end{align*}
It follows that 
$$
\|\Sigma - \left(R[0]-B_i A_i^{-1}B_i^\top \right)\|_2 \le \|H_{i,\mathrm{tail}}\|_{L_{\infty}}^2 \|\Phi\|_{L_{\infty}}.
$$
The bound then follows from Lemma~\ref{lem:predictorCoeffBound} and Assumption~\ref{assump:postive_psd}. 

\hfill $\qed$

\subsection{Proof of Theorem~\ref{thm:exact}}
Using \eqref{eq:conditionalMIBayes}
\begin{align*}
  \MoveEqLeft[0]
  I(\bx[0:i];\by[i]|\by[0:i],\bz[0:i])\\
  &=\bbE\left[
    \log\frac{p_{Y[i]|X[0:i]Y[0:i]Z[0:i]}(\by[i]|\bx[0:i],\by[0:i],\bz[0:i])}{
p_{Y[i]|Y[0:i]Z[0:i]}(\by[i]|\by[0:i],\bz[0:i])
    }
    \right].
\end{align*}
Let 
\begin{align*}
  \bbE[\bw[0:i+1]\bw[0:i+1]^\top] 
  =\begin{bmatrix}
    A_i & B_i^\top \\
    B_i & R[0]
  \end{bmatrix},
\end{align*}
where $A_i$ and $B_i$ were  defined in Lemma~\ref{lem:predictionCovBounds}.  
Bayes rule for Gaussians implies that conditioned on $\bw[0:i]$, $\bw[i]$ is a Gaussian random variable with mean $B_i A_i^{-1}\bw[0:i]$ and covariance $R[0]-B_i A_i^{-1} B_i^\top$.

Partition $R[0]$ and $B_i$ into blocks conforming to the sizes of $\bx$, $\by$, and $\bz$ as:
\begin{align*}
  R[0]&=\begin{bmatrix}
    R_{xx}[0] & R_{xy}[0] & R_{xz}[0] \\
    R_{yx}[0] & R_{yy}[0] & R_{yz}[0] \\
    R_{zx}[0] & R_{zy}[0] & R_{zz}[0]
  \end{bmatrix} 
  &
    B_i&=\begin{bmatrix}
    B_{i,x} \\
    B_{i,y} \\
    B_{i,z}
    \end{bmatrix}.
\end{align*}

Set $S_i = R_{yy}[0]-B_{i,y}A_i^{-1}B_{i,y}$.
Say that $\by[i]\in\bbR^{n_Y}$. 
Then, since $\bw[0:i]$ is precisely a re-arrangement of the entries of $\bx[0:i]$, $\by[0:i]$, and $\bz[0:i]$, we have that 
\begin{multline*}
  p_{Y[i]|X[:i]Y[:i]Z[:i]}(y[i]|x[:i],y[:i],z[:i])=\\
  \frac{1}{(2\pi)^{\frac{n_Y}{2}}\left(\det\left(S_i\right)\right)^{\frac{1}{2}}}
  \exp\left(-\frac{1}{2}
  (y[i]-B_{i,y}A_i^{-1}\bw[0:i])^\top S_i^{-1} \cdot\right. \\\left. (y[i]-B_{i,y}A_i^{-1}\bw[0:i])
  \right).
\end{multline*}

It follows that 
\begin{multline*}
  \bbE\left[
\log p_{Y[i]|X[0:i]Y[0:i]Z[0:i]}(\by[i]|\bx[0:i],\by[0:i],\bz[0:i])
\right]\\
=-\frac{n_Y}{2}\log(2\pi e)-\frac{1}{2}\log\det(S_i).
\end{multline*}

Now, let 
\begin{align*}
  C_i&=\begin{bmatrix}
    C_{i,y} \\
    C_{i,z}
  \end{bmatrix}=\bbE[\bv[i]\bv[0:i]^\top] \\
    D_i&=\bbE[\bv[0:i]\bv[0:i]^\top] \\
    U_i&=R_{yy}[0]-C_{i,y}D_i^{-1}C_{i,y}^\top.
\end{align*}

An argument analogous to the analysis of $\by[i]$ conditioned on $\bw[0:i]$ gives:
\begin{multline*}
  \bbE\left[
\log p_{Y[i]|Y[0:i]Z[0:i]}(\by[i]|\by[0:i],\bz[0:i])
\right]\\
=-\frac{n_Y}{2}\log(2\pi e)-\frac{1}{2}\log\det(U_i).
\end{multline*}

It follows that
$$
I(\bx[0:i];\by[i]| \by[0:i],\bz[0:i])=\frac{1}{2}\log\left(\frac{\det(U_i)}{\det(S_i)} \right).
$$

Lemma~\ref{lem:covConvergence} implies that $\lim_{i\to\infty}S_i = \Sigma_{yy}$. Similarly, $\lim_{i\to\infty}U_i = \Gamma_{yy}$. It follows that 
$$
\lim_{i\to\infty}
I(\bx[0:i];\by[i]| \by[0:i],\bz[0:i])=\frac{1}{2}\log\left(\frac{\det(\Gamma_{yy})}{\det(\Sigma_{yy})} \right).
$$
The proof is finished by noting that 
$$
  I^{\infty}(\bx\to\by\|\bz) = \lim_{k\to\infty}\frac{1}{k}\sum_{i=1}^{k-1} I(\bx[0:i];\by[i]| \by[0:i],\bz[0:i]).
$$
  \hfill$\qed$

\section{Proof of Theorem~\ref{thm:error}}
\label{sec:estimatorProof}

The estimation error in $\|\tilde\bR-\underline{R}\|_2$ is quantified, and then a collection of elementary matrix bounds are given. These preliminary results are then used to prove Theorem~\ref{thm:error}.

\subsection{Probabilistic Error Bound}

\begin{lemma} \label{lem:error_R_est2real}
  {\it Let $\tilde\bR$ be defined from \eqref{eq:empCov} and set $\underline{R}  =\bbE[\tilde\bR]$. Say that $\bw[k]\in\bbR^{n_W}$.  
    If Assumption \ref{assump:postive_psd} holds and $M>0$, then, for all $\nu \in (0,1)$, the following bound holds with probability at least $1-\nu$:
  \begin{align*}
    & \|\underline{R} - \tilde{\bR}\|_2 \le  \\  
    &2 c_{\max}\max\left\{\!\sqrt{(2p+1)8 \left(\frac{\log(2\nu^{-1})}{M} \!+ \!\frac{n_W(1+p)}{M} \log (13) \right)}, \right.  \\
    & \hspace{10pt}\left. (p+1)8 \left(\frac{\log(2\nu^{-1})}{M} + \frac{n_W(1+p)}{M} \log (13) \right) \right\}.
  \end{align*}
}
\end{lemma}

{\it Proof:}
Since $\tilde\bR-\underline{R}$ is symmetric, the variational characterization of the matrix $2$-norm implies that 
\begin{align*}
  \|\tilde{\bR} - \underline{R} \|_2 &= \max_{\|v\|_2 \le 1, s \in \{-1,1\}} s v^\top (\tilde{\bR} - \underline{R}) v.
\end{align*}
We will use a covering argument to reduce the problem of bounding the norm to bounding a finite number of terms of the form on the right. 

Let $\bbB$ denote the unit ball of $\bbR^{n_W+n_W p}$.
Let $\cC\subset\bbB$ be a $\delta$-cover of $\bbB$, i.e., $\forall \|u\|_2 \le 1$, $\exists v \in \cC$ s.t. $\|v\|_2 \le 1$ and $\| u - v \|_2 \le \delta$. As discussed in Example 5.8 of \cite{wainwright2019high}, the cardinality of $\cC$ is bounded by
$$
|\cC| \le \left(1 + \frac{2}{\delta}\right)^{n_W + n_W p}.
$$
For any vector, $u$, with $\|u\|_2\le 1$, there is a vector $v\in \cC$ such that 
$\|u-v\|_2 \le \delta$. Then, for $s\in \{-1,1\}$ we have
\begin{align*}
  s u^\top (\tilde{\bR} -\underline{R}) u &= s (v + u-v )^\top (\tilde{\bR} -\underline{R}) (v + u-v) \\
  & =  s v^\top (\tilde{\bR} -\underline{R}) v + s (u-v )^\top (\tilde{\bR} -\underline{R}) ( u-v) \\
  & +  s ( u-v )^\top (\tilde{\bR} -\underline{R}) v  +s v ^\top (\tilde{\bR} -R) (u-v) \\
  & \le  s v^\top (\tilde{\bR} -\underline{R}) v + \|\tilde{\bR} - \underline{R} \|_2 (2 \delta + \delta^2).
\end{align*} 

Maximizing both gives:
\begin{align*}
\sup_{(s,v) \subset \{-1,1\} \times \bbB} s v^\top (\tilde{\bR} -\underline{R}) v &\le \max_{(s,v) \subset \{-1,1\} \times \cC} s v^\top (\tilde{\bR} -\underline{R}) v \\
& + \|\tilde{\bR} - \underline{R}\|_2 (2 \delta +\delta^2).
\end{align*}
Therefore, if $2 \delta +\delta^2 <1$, i.e. $\delta < \sqrt{2} -1$, we have the following:
\begin{align*}
  \|\tilde{\bR} - \underline{R}\|_2 \le \frac{1}{1-3 \delta} \max_{(s,v) \subset \{-1,1\} \times \cC} s v^\top (\tilde{\bR} -\underline{R}) v.
\end{align*}
In order to obtain the bound with high probability of the left side, it suffices to bound $sv^\top (\tilde{\bR}-\underline{R})v$ for fixed $s$ and $v$ and then use a union bound. 

We will utilize an argument based on the Hanson-Wright inequality. 

Recall that $N=M+p$.  For $k=0,\ldots,M-1$, let $F_k$ be the matrix such that $\frac{1}{\sqrt{M}}\bw[k:k+p+1]=F_k\bw[0:N]$. 
Note that $F_k$ will have the form $F_k=\frac{1}{\sqrt{M}}\begin{bmatrix}0 & I & 0\end{bmatrix}$ for zero matrices of appropriate size. 

Let $\bU=\uwave{R}^{-1/2}\bw[0:N]$ where $\uwave{R} = \bbE[\bw[0:N] \bw[0:N]^\top]$. Then $\bU$ is a Gaussian vector with mean $0$ and covariance $I$. Then, we can express:
$$
\tilde\bR = \sum_{i=0}^{M-1}F_i \uwave{R}^{1/2} \bU\bU^\top \uwave{R}^{1/2}F_i^\top.
$$
If $A = s\uwave{R}^{1/2} \left( \sum_{k=0}^{M-1} F_k^\top v v^\top F_k \right) \uwave{R}^{1/2}$, then $sv^\top \tilde \bR v = \bU^\top A \bU$. Recall that $\bbE[\tilde \bR]=\underline{R}$. 

The Hanson-Wright inequality for Gaussians implies that 
\begin{equation} \label{eq:Gaussian_Hanson_Wright}
\begin{aligned}
&\bbP\left(sv^\top \tilde{\bR} v  - s v^\top \underline{R} v > \epsilon \right) \\
& \hspace{30pt} \le \exp\left( -\frac{1}{8} \min \left\{\frac{\epsilon^2}{\|A\|_F^2}, \frac{\epsilon}{\|A\|_2}\right\}\right) .
\end{aligned}
\end{equation}
See \citep{vershynin2018high} for a general discussion of the Hanson-Wright inequality and \citep{lamperski2023nonasymptotic} for a derivation of the explicit $1/8$ factor for Gaussians.

So, we need to bound the matrix norms. 

Note that
\begin{align*}
  \|A\|_F\le  \|\uwave{R}\|_2  \left\|  \sum_{k=0}^{M-1} F_k^\top v v^\top F_k \right\|_F \!.
\end{align*}

Lemma~\ref{lem:covBounds} implies that $\|\uwave{R}\|_2 \le c_{\max}$.

Then we calculate
\begin{align*}
  \left\|  \sum_{k=0}^{M-1} F_k^\top v v^\top F_k \right\|_F^2&=\sum_{i=0}^{M-1}\sum_{k=0}^{M-1}\Tr\left(F_i^\top vv^\top F_i F_k^\top vv^\top F_k \right) \\
                                                              &=\sum_{i=0}^{M-1}\sum_{k=0}^{M-1}\left(v^\top F_i F_k^\top v \right)^2. 
\end{align*}
To bound the right, note that $F_iF_k^\top$ can only be non-zero when the index sets $\{i,\ldots,i+p\}$ and $\{k,\ldots,k+p\}$ have a non-empty intersection. In turn, a non-empty intersection can only occur when $|i-k|\le p$. For each $i\in \{0,\ldots,M-1\}$, there are at most $2p+1$ values of $k\in \{0,\ldots,M-1\}$ such that $F_i F_k^\top\ne 0$. For any such pair, 
$$
\left(v^\top F_i F_k^\top v\right)^2\le \|F_i^\top v\|_2^2 \|F_k^\top v\|_2^2 = \frac{\|v\|_2^4}{M^2}.
$$
Using that $\|v\|_2\le 1$, it follows that 
$$
\left\|  \sum_{k=0}^{M-1} F_k^\top v v^\top F_k \right\|_F^2\le \frac{2p+1}{M}\implies 
\|A\|_F^2\le c_{\max}^2 \frac{2p+1}{M}.
$$

To bound $\|A\|_2$, note that 
$$
\|A\|_2\le \|\uwave{R}\|_2 \left\|\sum_{k=0}^{M-1}F_k v v^\top F_k^\top \right\|_2.
$$
Then, since $\|v\|_2\le 1$, we have
\begin{align*}
  \MoveEqLeft
  \left\|  \sum_{k=0}^{M-1} F_k^\top v v^\top F_k \right\|_2 \\&= \max_{\|u\|_2 \le 1} u^\top \left(\sum_{k=0}^{M-1} F_k^\top v v^\top F_k  \right) u  \\
                                                               & =\frac{1}{M}\max_{\|u\|_2\le 1}\sum_{i=0}^{M-1}(u[i:i+p+1]^\top v)^2
\\
                                                               &\le \frac{1}{M}\max_{\|u\|_2\le 1}\sum_{i=0}^{M-1}\|u[i:i+p+1]\|_2^2\\
                                                               &\le \frac{p+1}{M}.
\end{align*}
The final inequality follows because each term, $\|u[i]\|_2^2$,  appears in the sum at most $p+1$ times. 

Plugging the bounds on $\|A\|_2$ and $\|A\|_F$ into \eqref{eq:Gaussian_Hanson_Wright} gives
\begin{align*}
  &\bbP\left(sv^\top \tilde{\bR} v  - s v^\top \underline{R} v > \epsilon \right) \\
  & \hspace{30pt} \le \exp\left( -\frac{1}{8} \min \left\{\frac{\epsilon^2}{c_{\max}^2 \frac{2p+1}{M}}, \frac{\epsilon}{c_{\max} \frac{p+1}{M}}\right\}\right) \\
  &  \hspace{30pt} \le \exp\left( -M \min \left\{\frac{\epsilon^2}{c_{\max}^2 (2p+1)8}, \frac{\epsilon}{c_{\max} (p+1) 8 }\right\}\right) .
\end{align*}

Set $\kappa(\epsilon) = \min \left\{\frac{\epsilon^2}{c_{\max}^2 (2p+1)8}, \frac{\epsilon}{c_{\max} (p+1) 8 }\right\}$. Note that $\kappa$ is invertible mapping from $[0,\infty)$ to $[0,\infty)$, with 
$$
\kappa^{-1}(y)=c_{\max}\max\left\{\sqrt{(2p+1)8 y},(p+1)8 y\right\}.
$$

Then, we can obtain the following union bound:
\begin{align*}
  &\bbP\left( \|\underline{R} - \tilde{\bR}\|_2 > \epsilon \right) \\
  &\le \sum_{(s,v) \in \{-1,1\}\times \cC } \bbP\left( sv^\top (\underline{R} -\tilde{\bR})  v  > (1- 3 \delta) \epsilon\right) \\
  & \le 2 \left(1+\frac{2}{\delta}\right)^{n_W+n_W p} \exp(-M\kappa\left((1- 3 \delta) \epsilon)\right).
\end{align*}
Set the right side to $\nu$, for $\nu\in (0,1)$. Then, we have the following:
\begin{align*}
\log(\nu)\! =\! \log 2 \!+\! (n_W+n_W p) \log \left(1+ \frac{2}{\delta}\right) \!-\! M \kappa\left((1- 3 \delta) \epsilon\right).
\end{align*}
For simplicity set $\delta = \frac{1}{6}$ above, to give 
  \begin{align*}
    \kappa\left(\frac{1}{2} \epsilon\right) = \frac{\log(2\nu^{-1})}{M} + \frac{n_W+n_Wp}{M} \log (13)  .
    \end{align*}

The result follows by solving for $\epsilon$.
 \hfill $\qed$

\subsection{Elementary Matrix Calculus}

\begin{lemma}
  \label{lem:logDetDiff}
  {\it
Let $A$ and $B$ be positive definite $n\times n$ matrices. 
\begin{multline*}
\left|\log\det(A)-\log\det(B) \right|\le \\n\max\{\|A^{-1}\|_2,\|B^{-1}\|_2\} \|A-B\|_2.
\end{multline*}
}
\end{lemma}

{\it Proof:}
Let $f(t)=\log\det(A+t(B-A))$, so a standard matrix calculation, e.g. Section 0.8.10 of \cite{horn2012matrix}, gives that 
$$
\frac{df(t)}{dt}=\Tr\left((A+t(B-A))^{-1} (B-A) \right).
$$
The mean value theorem implies that there is some $t\in [0,1]$ such that 
$$
\log\det(B)-\log\det(A)=\frac{df(t)}{dt}.
$$

The von Neuman trace inequality implies that 
\begin{multline*}
\left|\Tr\left((A+t(B-A))^{-1} (B-A) \right)\right| \le \\ n \| (A+t(B-A))^{-1}\|_2 \|A-B\|_2.
\end{multline*}

The function $g(A)=\|A^{-1}\|_2$ is convex over the set of positive definite matrices. Indeed, $(A,\alpha)$ is in its epigraph if and only if 
$$
\begin{bmatrix}
  \alpha I & I \\
  I & A
\end{bmatrix} \succeq 0.
$$
Convexity, and the fact that $t\in [0,1]$ implies that 
\begin{align*}
  \| (A+t(B-A))^{-1}\|_2 &\le (1-t) \|A^{-1}\|_2 + t\|B^{-1}\|_2 \\
                         &\le \max\{\|A^{-1}\|_2,\|B^{-1}\|_2\}.
\end{align*}
Combining the bounds gives the result. 
\hfill$\qed$

\begin{lemma}
  \label{lem:schurInvBound}
  {\it
  If 
  $$
  \begin{bmatrix}
    A & B^\top \\
    B & C 
  \end{bmatrix} \succeq cI 
  $$
  for some $c>0$,
  then the Schur complement satisfies
  $$
  \left\|\left(C-B A^{-1}B^\top\right)^{-1}\right\|_2\le c^{-1}.
  $$
}
\end{lemma}

{\it Proof:}
The following direct calculation suffices:
\begin{align*}
  C-B A^{-1}B^{\top} &=\begin{bmatrix}
    -BA^{-1} & I
  \end{bmatrix}
  \begin{bmatrix}
  A & B^\top \\
  B & C
  \end{bmatrix}
  \begin{bmatrix}
    -A^{-1}B^\top \\
  I
  \end{bmatrix} \\
                   &\succeq c \begin{bmatrix}
                     -BA^{-1} & I
  \end{bmatrix}
  \begin{bmatrix}
    -A^{-1}B^\top \\
  I
  \end{bmatrix}  \\
                   &\succeq cI.
\end{align*}
\hfill$\qed$

\begin{lemma}
  \label{lem:schurDiff}
  {\it
  If the following bounds hold 
  \begin{gather*}
    c_1 I \preceq \begin{bmatrix}A & B^\top \\
    B & C \end{bmatrix} \preceq c_2 I,  \quad
      c_1 I \preceq \begin{bmatrix}\hat A & \hat B^\top \\
    \hat B & \hat C \end{bmatrix} \preceq c_2 I\\
    \left\|
 \begin{bmatrix}A & B^\top \\
    B & C \end{bmatrix} -
 \begin{bmatrix}\hat A & \hat B^\top \\
    \hat B & \hat C \end{bmatrix} 
    \right\|_2\le \delta
  \end{gather*} 
  for some $c_1 >0$, $c_2 >0$, and $\delta >0$, then
  $$
  \left\|\left(C-B A^{-1}B^\top \right) - \left(
  \hat C-\hat B \hat A^{-1}\hat B^\top
\right)\right\|_2 \le \delta\left(1+\frac{c_2^2}{c_1^2}\right).
  $$
}
\end{lemma}
{\it Proof:}
Applying Lemma~\ref{lem:completeSquare} with $\Theta=\hat B \hat A^{-1}$ implies that 
\begin{align*}
\MoveEqLeft
\left(C-B A^{-1}B^\top \right) - \left(
  \hat C-\hat B \hat A^{-1}\hat B^\top
\right) \\
&\preceq 
\begin{bmatrix}
  -\hat B \hat A^{-1} & I
\end{bmatrix}
\left( 
  \begin{bmatrix}
  A & B^\top \\
  B & C
  \end{bmatrix} -
  \begin{bmatrix}
  \hat A & \hat B^\top \\
  \hat B & \hat C
  \end{bmatrix}
    \right)
\begin{bmatrix}
  -\hat A^{-1}\hat B^\top \\
  I
  \end{bmatrix}\\
&\preceq \delta 
\begin{bmatrix}
  -\hat B \hat A^{-1} & I
\end{bmatrix}
\begin{bmatrix}
  -\hat A^{-1}\hat B^\top \\
  I
  \end{bmatrix}\\
&\preceq \delta \left(1+\frac{c_2^2}{c_1^2} \right)I.
\end{align*}
The final bound follows because $\|\hat B\|_2\le c_2$ and $\|\hat A^{-1}\|_2 \le c_1^{-1}$. 

A  lower bound of $-\delta \left(1+\frac{c_2^2}{c_1^2}\right)$ is derived similarly.
\hfill$\qed$

\subsection{Proof of Theorem~\ref{thm:error}}
First, we give a sufficient condition for the required inverses to exist. 
For compact notation, let $\epsilon$ denote the upper bound on $\|\tilde\bR-\underline{R}\|_2$ from Lemma~\ref{lem:error_R_est2real}.

Then, with probability at least $1-\nu$:
$$
(c_{\min}-\epsilon)I\preceq \tilde \bR \preceq (c_{\max} + \epsilon) I.
$$
Note that $\epsilon=O\left(\frac{p\log(\nu^{-1})}{M}\right)$. So $\epsilon < c_{\min}$ for sufficiently large $M/p$, and in this case $\tilde \bR$ is positive definite with probability at least $1-\nu$.  

To bound $|I^{\infty}(\bx\to\by\|\bz)-\tilde\bI|$, it suffices to bound:
\begin{gather*}
  \left| \log\det(\Gamma_{yy})-\log\det(\tilde \bGamma_{yy})\right| \quad \textrm{and} \\
\left| \log\det(\Sigma_{yy})-\log\det(\tilde \bSigma_{yy})\right|.
\end{gather*}
We focus on bounding $\left| \log\det(\Sigma_{yy})-\log\det(\tilde \bSigma_{yy})\right|$, since the other term will satisfy the same bound. 

If $\by[k]\in\bbR^{n_Y}$, then Lemma~\ref{lem:logDetDiff} implies that
\begin{multline*}
\left| \log\det(\Sigma_{yy})-\log\det(\tilde \bSigma_{yy})\right|\\\le n_Y \max\{\|\Sigma_{yy}^{-1}\|_2,\|\tilde\bSigma_{yy}^{-1}\|_2\}\|\Sigma_{yy}-\tilde\bSigma_{yy}\|_2.
\end{multline*}

Lemma~\ref{lem:schurInvBound} implies $\|\Sigma_{yy}^{-1}\|_2\le c_{\min}^{-1}$. Similarly, using the lower bound on $\tilde\bR$ and Lemma~\ref{lem:schurInvBound} implies that $\|\tilde\bSigma_{yy}^{-1}\|_2\le \frac{1}{c_{\min}-\epsilon}$. 

Let $S_p=R_{yy}[0]-B_{p,y}A_p^{-1}B_{p,y}^\top$. The triangle inequality, followed by Lemma~\ref{lem:covConvergence} implies that:
\begin{align*}
  \|\Sigma_{yy}-\tilde\bSigma_{yy}\|_2 &\le   \|\Sigma_{yy}-S_p\|_2 + \|S_p-\tilde\bSigma_{yy}\|_2\\
                                       &\le c_{\max}\frac{b^2 \rho^{2(p+1)}}{(1-\rho)^2}+\|S_p-\tilde\bSigma_{yy}\|_2.
\end{align*}

Then Lemma~\ref{lem:schurDiff} implies that 
$$
\|S_p-\tilde\bSigma_{yy}\|_2\le \epsilon\left(1+\frac{(c_{\max}+\epsilon)^2}{(c_{\min}-\epsilon)^2} \right).
$$

Putting the bounds together gives:
\begin{multline*}
  \left| \log\det(\Sigma_{yy})-\log\det(\tilde \bSigma_{yy})\right| \le 
  \frac{n_Y}{c_{\min}-\epsilon}\cdot \\
\left(
c_{\max}\frac{b^2 \rho^{2(p+1)}}{(1-\rho)^2}+
\epsilon\left(1+\frac{(c_{\max}+\epsilon)^2}{(c_{\min}-\epsilon)^2} \right)
\right).
\end{multline*}

Now, $\left|\log\det(\Gamma_{yy})-\log\det(\tilde\bGamma_{yy})\right|$ satisfies the same bound, and so the factor of $1/2$ cancels out, and the main bound has been proved.

In the case that $p\ge \frac{\log(N)}{4\log(\rho^{-1})}$, we have that $\rho^{2p}\le N^{-1/2}$. Furthermore, if $p\le a_2 \log(N)$, we have that $M=N-p=\Omega(N)$. Plugging the bounds on $M$ and $p$ into the expression for $\epsilon$ shows that $\epsilon=O\left(\log(N)\sqrt{\frac{\log(\nu^{-1})}{N}}\right)$. Plugging the bound on $\epsilon$ into the bound above gives the result.
\hfill$\qed$

\section{Simulation} \label{sec:simulation}

In this section, we present the errors, $|I^\infty(\bx\to\by\| \bz)-\tilde\bI|$, for a VAR(1) model as below:
\begin{align*}
\bs[k+1] &= A \bs[k] + B \be[k]
\end{align*}
where 
 $\be \sim \cN(0, I)$.

And the measurements are 
\begin{align*}
\bw[k] &= C_w \bs[k] + D_w \be[k] \\
\bv[k] &= C_v \bs[k] + D_v \be[k]. \\
\end{align*}
More details on the simulation are available on GitHub\footnote{\url{https://github.com/zhen0348/errorBounds_CCDI}}.

\begin{figure}[H]
      \centering
      \includegraphics[width=.7\columnwidth]{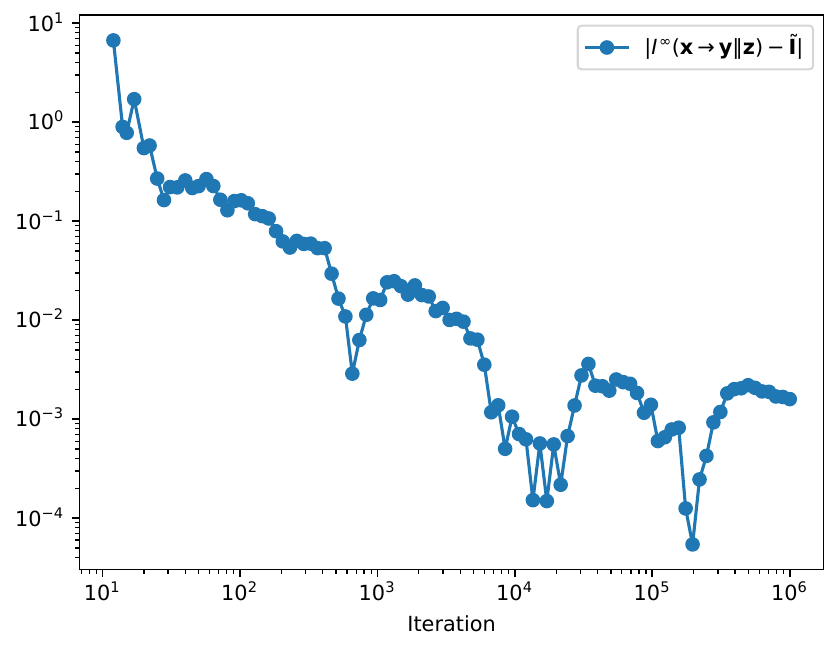}
	\caption{Errors between theoretical causally conditioned DI rates and the empirical values in the VAR(1) simulation}
	\label{fig:errors}
  \end{figure}

Note we use Kalman filter to to calculate the theoretical causally conditioned DI rates. Fig. \ref{fig:errors} shows that the errors converge to a small value. Here, the theoretical bound in Theorem \ref{thm:error} is not shown because it is rather conservative. Getting tighter error bounds remains to be studied in the future.

\section{Conclusion and Future Work}
\label{sec:conclusion}

This paper presents a non-asymptotic error bound for the causally conditioned DI rate for stationary Gaussian sequences. The algorithm was enabled by a simple formula for the causally  conditioned DI rate in terms of prediction error covariance matrices. 
Future work will focus on non-Gaussian and possibly non-stationary sequences. Such extensions will be applicable for the analysis of more complex dynamical systems, with nonlinear and time-varying behaviors. 


\bibliography{cool-refs}



\end{document}